\documentstyle[epsfig,12pt]{article}
\newcommand{\rar}{\rightarrow}

\newcommand{\be}{\begin{eqnarray}}
\newcommand{\ee}{\end{eqnarray}}
\begin{document}

\begin{center}
{\large\bf
K-MATRIX ANALYSIS OF THE $K \pi$ S-WAVE IN THE MASS REGION
$900-2100$ MEV AND NONET CLASSIFICATION OF SCALAR $q\bar q$-STATES}
\end{center}

\vspace{1.0cm}

\centerline{A.V. Anisovich and A.V. Sarantsev}
\vspace{0.5cm}
\centerline{Petersburg Nuclear Physics Institute}
\centerline{188350 Gatchina, Russia}
\vspace{0.5cm}

\begin{abstract}

A K-matrix
re-analysis of the $K\pi$ S-wave is performed
in the mass region $900-2100$
MeV, and  solutions which describe  the data well are
found. The solution with two resonances in the
$(IJ^P=\frac{1}{2}0^+)$-wave has poles at
$(1415 \pm 25 )$- $i(165 \pm 25 )$ MeV and
$(1820 \pm 40 )$ -$i(125 \pm 50 )$ MeV, close to the previous
result of D. Aston et al., Nucl.Phys., B296 (1988) 493. The
corresponding bare states, which are the subjects of a $q\bar q$
classification, are: $K_0^{bare}(1220^{+50}_{-60})$ and
$K_0^{bare}(1885^{+50}_{-80})$.
Within the restored bare states, a construction of the
 $1^3P_0\; q\bar q$ and $2^3P_0\; q\bar q$ nonets is completed.
The three resonance 
solutions are analysed as well, the corresponding
 bare states are found, and the $q\bar q$-nonet classification
 is suggested.
 In all variants of our $q\bar q$-classification, the resonances $a_0(980)$
and $f_0(980)$ (or their bare counterparts) are non-exotic, 
 and an extra scalar/isoscalar state exists
in the mass region 1300-1700 MeV, being a good candidate for the
lightest scalar glueball.

\end{abstract}

\section{Introduction}
The strategy for searching for exotic mesons,
which is outlined in our recent investigations is based on a
systematic  classification of $q\bar q$ meson states: the extra states
for such classification are to be considered as candidates for exotics
(glueballs, hybrids, etc).

However, a systematization of the $q\bar q$ states at
masses above 1000 MeV faces a serious problem: many of the observed
resonances are products of the mixing caused by transitions $q\bar q\;
state \to real\; mesons \to q\bar q\; state$ at large distances,
$r>R_{confinement}$.  Correspondingly, the wave functions of the mixed
states (resonances) contain the large-$r$ multi-meson components with
significant probabilities. The wave functions
 restored in such a way 
cannot be compared with the results of Strong-QCD models, which do not
take into account the deconfinement of the quarks due to the
resonance decay.  For a comparision of the results of data-analysis  with
quark model calculations and for a classification of  $q\bar q$-levels,
one needs to separate large-$r$ and small-$r$ wave function components.

In refs. \cite{km1500,km1900}, it was suggested to carry out the $q\bar q$
classification in terms of states which correspond to the K-matrix
poles (bare states): their relation to the input poles of the
propagator matrix (or D-matrix) is discussed in detail in ref.\cite{aas}.

The K-matrix analysis of the ($IJ^{PC}$=$00^{++}$)-wave which
has been performed in the
mass region 600-1950 MeV \cite{km1900} points out an existence of five
meson states: four of them should be considered as candidates for the
members of two lightest nonets, $1^3P_0 q\bar q$ and $2^3P_0 q\bar q$,
while the fifth state with the mass in the region 1250-1650 MeV is a
good candidate for the scalar glueball. More detailed analysis of this
wave carried out in terms of  D-matrix and $q\bar q$
transition amplitudes \cite{aas,akm1900} confirms the results of
ref.\cite{km1900}, restoring a quark/gluonium content of the observed
resonances as well as the masses of non-mixed states (in
particular, the mass of the pure scalar glueball which is the subject of
the gluodynamic lattice calculations \cite{ukqcd}).

Recently the K-matrix analysis was extended to the $IJ^{PC}=10^{++}$ wave
at  600-1800 MeV \cite{fullkm} thus allowing us to determine
$a_0^{bare}$'s in this mass region. To a complete
the construction of
$1^3P_0q\bar q$ and $2^3P_0q\bar q$ nonets in terms of bare states, the
K-matrix analysis of the $IJ^{PC}=\frac{1}{2}0^{++}$ state has to be
performed:  only after that one can definetely declare an exisance of
extra states for the $q\bar q$ classification.

The classification of $\frac{1}{2}0^+$-resonances faces a problem:
the mass of the lightest scalar K-meson, $1429\pm9$ MeV \cite{Aston88},
is significantly higher then the masses the other lightest scalars,
$f_0(980)$ and $a_0(980)$. 
In the set of papers (see, for example, \cite{exotics})
the resonances
$f_0(980)$ and $a_0(980)$ were considered 
as superfluous for the
$q\bar q$ classification and the lightest scalar $q\bar
q$ nonet was built from the higher resonances.
 Therefore, it is of key importance in a search for
exotic mesons to perform the K-matrix analysis of the
$\frac{1}{2}0^{++}$ wave. The present paper is devoted to this
problem.

Let us stress that the
identification of scalar/isoscalar resonances below 1950 MeV
is quite reliable now because of  significant progress
during the few last years (see refs.
\cite{km1900,cbar,bugg}
and references therein).
The simultaneous analysis of Crystal Barrel \cite{cbar},
CERN-M\"unich \cite{cern}, GAMS \cite{gams},
and BNL \cite{bnl} meson
spectra performed in ref. \cite{km1900} shows that
the $IJ^{PC}=00^{++}$ amplitude has five
poles below 1900 MeV located at (in MeV units):
$\;(1015\pm 15)- i(43\pm 8)$, $\;(1300\pm 20)-i(120\pm 20)$,
$\;(1499\pm 8)-i(65\pm 10)$, $\;(1780\pm 30)-i(125\pm 70)$,
 $\;(1530^{+90}_{-250}) - i(560\pm 140)$.
The first four poles correspond to comparatively narrow
resonances $f_0(980)$, $f_0(1300)$, $f_0(1500)$ and $f_0(1780)$;
the fifth pole is related to a broad state defined
in \cite{km1900} as $f_0(1530^{+90}_{-250})$.
The states $f_0(980)$ and $f_0(1500)$ are included now in the
list of well defined resonances \cite{pdg}. The properties of
two states $f_0(1300)$ and $f_0(1530^{+90}_{-250})$ are collected
together and defined as $f_0(1370)$ resonance in \cite{pdg}.
The resonance $f_0(1780)$
reproduces well
the  bump at 1750 MeV in $\pi\pi\to K\bar K$ data \cite{bnl}
and is partly responsible for the bump in the GAMS data on
$\pi\pi\to\eta \eta'$  \cite{gams}.

Correspondingly, the five bare states are unambiguously
defined in \cite{km1900}:
\be
f^{bare}_0(720\pm 100),\; f^{bare}_0(1235^{+150}_{-30}),\;
f^{bare}_0(1260^{+100}_{-30}),\; f^{bare}_0(1600\pm 50),\;
f^{bare}_0(1810^{+30}_{-100}).
\label{1}
\ee
However, the $q\bar q$ classification of the bare scalar/isoscalar
states faces uncertainties:
there are two variants of the the construction
of the $1^3P_0q\bar q$ and $2^3P_0q\bar q$ nonets
related to a different choice of the bare state as the scalar gluonium.
The source of such ambiguity  
 is that the coupling constants for the gluonium decay
coincide with that
for transitions $q\bar q\; bare\; state\to two\; pseudoscalar\;
mesons$
at some definite $n\bar n / s\bar s $ content
(here $n\bar n=(u\bar u+d\bar d)/\sqrt{2}$): namely, at
$\phi=25^{\circ} -30^{\circ}$ where $\phi $ is determined as
 $q\bar q=n\bar n\; \cos\phi +s\bar s\; \sin\phi $.

It results in two classification
schemes  \cite{km1900}:
\begin{description}
\item[I.~~~] $f^{bare}_0(720)$ and $f^{bare}_0(1260)$ are members of
the $1^3P_0q\bar q$ nonet;\\
 $f^{bare}_0(1600)$ and $f^{bare}_0(1810)$
are members of $2^3P_0q\bar q$ nonet;\\ $f^{bare}_0(1235)$ is a
glueball.
 \item[II.~~~]$f^{bare}_0(720)$ and $f^{bare}_0(1260)$ are
members of the $1^3P_0q\bar q$ nonet; \\ $f^{bare}_0(1235)$ and
 $f^{bare}_0(1810)$ are members of the first radial excitation nonet
$2^3P_0q\bar q$;\\ $f^{bare}_0(1600)$ is a glueball.
\end{description}
Such ambiguity in the classification is
related to the close mixing angles and coupling constants for
$f^{bare}_0(1230)$ and $f^{bare}_0(1600)$ states.

K-matrix analysis of
isoscalar/tensor, isovector/scalar, and isovector/tensor amplitudes
is performed recently in ref.  \cite{fullkm}. The K-matrix approach,
describing well the data, gives the following $10^{++}$ bare states:
\be
a_0^{bare}(960\pm35),\; a_0^{bare}(1640\pm 40)\; .
\label{2}
\ee
The partial wave analysis of the $K^- \pi^+$ system for
the reaction $K^- p \rar K^- \pi^+ n$ at
11 GeV/c was performed by D.Aston et al. \cite{Aston88}.
Two alternative solutions for the S-wave partial amplitude were obtained
(following ref.\cite{Aston88} we call them solutions A and B); these
solutions differ only at masses above 1800 MeV. In ref.
\cite{Aston88}, the analysis of the $K\pi$ S-wave was made
separately for the mass regions $850-1600$ MeV and $1800-2100$ MeV,
where the isospin 1/2 partial wave amplitude was parametrized as the
sum of a Breit-Wigner resonance and a background term:
 \be
A^{1/2}_S=\sin\delta e^{i\delta} +
\frac{M_R\Gamma_1} {(M_R^2-M^2)-iM_R(\Gamma_1+\Gamma_2)}\;.
\label{3}
\ee
Here $\Gamma_1$ and $\Gamma_2$ are partial widths for the resonance
decay into $K \pi$ and $K \eta'$ channels.  In the first
mass region the
resonance $K^*_0(1430)$ was found:
\be
M_R=1429 \pm 9\; \mbox{MeV},
\;\;\; \Gamma= \Gamma_1 +\Gamma_2= 287 \pm 31 \; \mbox{MeV}\;.
\label{4}
\ee
In
the second mass region, two solutions A and B give the following
parameters for $K^*_0(1950)$:
  \be \mbox{Solution A:}\qquad\qquad
M_R=1934 \pm 28\; \mbox{MeV}, \;\;\;
\Gamma=  174 \pm 98 \; \mbox{MeV}.
\label{5}
\ee
\be
\mbox{Solution B:}\qquad\qquad
M_R=1955 \pm 18\; \mbox{MeV}, \;\;\;
\Gamma=  228 \pm 56 \; \mbox{MeV}.
\ee

However we believe that the analysis described above
should be extended. First of all, the mass region
$1600-1800$ MeV, where the amplitude varies rapidly, should not be
excluded from the analysis. It is known from the analysis
of the $00^{++}$ wave  that, due to strong
interference, the
resonance can manifest itself not only as a peak in the
spectrum but also as a dip or a shoulder: that happens
for $f_0(980)$ and $f_0(1500)$ in the $\pi\pi$ scattering
amplitude.
Secondly,
interference effects result in ambiguities. As a reminder,
 the ambiguities
in $00^{++}$-wave were resolved
in \cite{km1900} only due to a simultaneous fit of
a number of different meson spectra; but there is no such variety 
of data for the $\frac{1}{2}0^+$ wave. So, one can expect
that the solution found in \cite{Aston88} is not unique.

Concluding, the aim of the present investigation is:

$(i)$ To restore the masses and the meson decay couplings
of the scalar/isospin-$\frac{1}{2}$ bare states to perform a $q\bar
q$-classification;

$(ii)$ To obtain alternative K-matrix solutions for the
$K\pi$ S-wave in the region below 2000 MeV.

\section{K-matrix approach to the $K\pi$ S-wave amplitude.}

The S-wave $K^- \pi^0$ scattering amplitude extracted from the
$K^- p\to K^-\pi^+ n$ reaction at small momentum transfer
is a sum of the isospin-$\frac{1}{2}$ and isospin-$\frac{3}{2}$
components:
 \be
 A_S= A^{1/2}_S + A^{3/2}_S =\mid A_S
\mid e^{i\phi_S}
\label{6}
\ee
where $\mid A_S \mid$ and $\phi_S$  are  measured magnitudes
of the S-wave amplitude.

The isospin $I=3/2$ component of the $K \pi$ S-wave amplitude
is assumed to have non-resonant behaviour, so we use the following
parameterization:
\be
A_S^{3/2}(s)= \frac{\rho_1(s) a_{3/2}(s)}{1- i \rho_1(s)
a_{3/2}(s)},
\label{7}
\ee
where $a_{3/2}(s)$ is written as
\be
a_{3/2}(s)= a_{3/2}+ \frac{f_{3/2}}{s-s_{3/2}},
\label{8}
\ee
and $\rho_1(s)$ is the $K\pi$ phase volume:
\be
\rho_1(s)=\sqrt{\biggl(1-\frac{(m_K+m_\pi)^2}{s}\biggr)
                \biggl(1-\frac{(m_K-m_\pi)^2}{s}\biggr)}\;.
\label{9}
\ee

For the description of the $A^{1/2}_S$ amplitude, we
use the standard K-matrix representation:
\be
A^{1/2}_S=  K_{1a}\;(I-i\rho K)^{-1}_{a1}
\label{10}
\ee
where $K_{ab}$ is a 3 x 3 matrix
(a,b=1,2,3) with the following notations for meson
channels:
\be
1=K \pi, \;2=K\eta',\; 3=K \pi \pi \pi+ multimeson\;  states\; .
\label{11}
\ee
Including  the resonance decay into the $K \eta$ channel does
not affect the description  of the data:
the $K \eta$-coupling constant predicted by
quark combinatoric rules is comparatively small and agrees well with
experimental data (see \cite{Aston88} and references therein).
Thus,  below we give the solutions where the $K \eta$-channel is omitted.

The phase
space matrix is diagonal: $\rho_{ab}=\delta_{ab} \rho_a$ with
$\rho_1$ defined by eq.(\ref{9}) and $\rho_2$ is equal to:
\be
\rho_2(s)=\sqrt{\biggl(1-\frac{(m_K+m_{\eta'})^2}{s}\biggr)
                \biggl(1-\frac{(m_K-m_{\eta'})^2}{s}\biggr)}
\label{11a}
\ee
Below $K\eta'$ threshold, we use the analytic continuation
of the phase space factor:

$\rho_2= i\mid\rho_2\mid$.

For the $K\pi \pi \pi +multimeson$
phase space factor at $s< 1.44$ GeV$^2$, we have used three
alternative variants: it is defined either as
    \be
\rho_3(s)=C_1 \biggl( \frac{s-(m_K-3m_\pi)^2}{s} \biggr)^{5/2},
\; \; s<1.44\; \mbox{GeV}^2;
\label{rho1}
\ee
$$
\rho_3(s)=1 , \;\; s>1.44\; \mbox{GeV}^2,
 $$
or as   $\rho (770) K^*(892)$ and  $\sigma K^*(892)$ phase space
factors  ($\sigma$ stands for the
 low energy $\pi\pi$ S-wave amplitude):
\be
\rho_3(s)=C_2 \int
\limits_{(m_K+m_\pi)^2}^{(\sqrt{s}-2m_\pi)^2} dm^2_1
\int \limits_{4m_\pi^2}^{(\sqrt{s}-m_1)^2} dm^2_2
\frac{M_1 M_2 \Gamma_1 \Gamma_2}
{((M_1^2-m_1^2)^2+M_1^2\Gamma_1^2)
 ((M_2^2-m_2^2)^2+M_2^2\Gamma_2^2)}
   \nonumber \\
\times
 \sqrt{\biggl(1-\frac{(m_1+m_2)^2}{s}\biggr)
      \biggl(1-\frac{(m_1-m_2)^2}{s}\biggr)}, \;\;
 s<1.44\; \mbox{GeV}^2;
\label{rho2}
\ee
$$
\rho_3(s)=1 , \;\; s>1.44\; \mbox{GeV}^2.
 $$

Here $M_a$ and $\Gamma_a$ refer to the masses and widths of $K^*(892)$
and $\rho(770)$ (or $K^*(892)$ and $\sigma$). The factors $C_i$
provide continuity of $\rho_3(s)$ at $s=1.44$ GeV$^2$.
 The obtained characteristics of the $\frac{1}{2}0^+$-states 
 do not depend significantly on the form used for $\rho_3(s)$:
below we present results for $\rho_3(s)$ given by eq.(\ref{rho1}).

For $K_{ab}$, we use the following parameterization:
\be
K_{ab}(s)= \sum_{\alpha} \frac{g_a^{\alpha}
g_b^{\alpha}}{M^2_\alpha -s}\;+\; f_{ab}
\frac{1.5\; \mbox{GeV}^2-s_0}{s-s_0}\; .
\label{12}
\ee
The $g_a^{\alpha}$ are couplings of the bare states; the
 parameters $f_{ab}$ and
$s_0$ describe a smooth part (background) of K-matrix elements
($s_0 < 0$).

The $K^- p\to K^- \pi^+ n$ data were collected
at small momenta transfer, ($|t|<0.2$ GeV$^2$),
so as a first step we have used to fit the
data the unitary amplitude (\ref{10}). As the next step, we
have introduced a t-dependence into the K-matrix amplitude. For the
amplitude  $K \pi (t)\to K\pi$ (where $\pi (t)$ refers to a
virtual pion) we write:
\be
A^{1/2}_S=  \tilde K_{1a}\;(I-i\rho
K)^{-1}_{a1},
\label{13}
\ee
where
\be
\tilde K_{1a}(s)= \sum_{\alpha}
\frac{g^{\alpha}(t) g_a^{\alpha}}{M^2_\alpha -s}\;+\; f_{a}(t)
\frac{1.5\; \mbox{GeV}^2-s_0}{s-s_0}.
\label{b1}
\ee
In the limit $t\to m_\pi^2$, the
couplings $g^{\alpha}(t)$ and background terms $f_{a}(t)$
coincide with $g^{\alpha}_1$ and $f_{1a}$ correspondingly.
Taking into account that the momenta transfered are comparatively
small, $|t|<0.2$, we approximate  $g^{\alpha}(t)$ and $f_{a}(t)$
as
 \be
g^{\alpha}(t)= u^\alpha
g_1^\alpha,\qquad f_a(t)=u_a f_{1a}
\label{14}
\ee
where parameters
$u^\alpha$ and $u_a$ vary in the interval $0.9-1.1$.

In the leading terms of the $1/N$ expansion \cite{thooft}
, the couplings of the $q
\bar q$-meson  transition
to two mesons are determined by the diagram shown
in fig.1a where gluons produce a $q \bar q$-pair. The production
of soft $q \bar q$ pairs by gluons violates flavour symmetry
(see, for example, \cite{km1900,Anis90,Anis95}): the quark
production probability ratios are $u \bar u: d \bar
d: s \bar s=1:1:\lambda$ with $\lambda=0.45-0.8$.
We fix in our fit  $\lambda=0.6$.
This makes it possible to calculate the
decay coupling constants in the framework of the
quark combinatoric rules: they are given in Table 1 for both
the leading terms of the $1/N$-expansion (process of the
fig. 1a type) and for the next-to-leading terms (process of the fig 1b
type). In the present fit, we have used the leading terms only.

\section{Fit of the data and the $q\bar q$ classification schemes}

In ref. \cite{Aston88} there are two solutions, A and B, for the
$\frac120^+$ wave, which differ from each other at $M_{\pi K}>1800$ MeV
only. Correspondingly, we
have got two K-matrix two-pole solutions, A-1 and
B-1. In addition, for an analysis
of alternatives,  we have performed
the K-matrix three-pole fits.

The solid curves in figs. 2-6 correspond to the description of
the $K \pi$ wave defined by the unitary expression
(\ref{10}).  The dashed curves show fits where
t-dependence of the  $K\pi$ interaction
is taken into account by means of eq.(\ref{13}). As
it is seen, this dependence
 helps to describe phase shifts in the region
1700 MeV.
Let us note that in this region (and in the region higher 2000 MeV
in the solution A) the data violate unitarity limits. Such a strong
violation is unlikely to be expected by t-dependence alone: we think 
it is connected with inaccurate PW analysis in this region performed
in ref.\cite{Aston88}.
  However including t-dependence in our fits does
not produce any serious effect either on  masses of the
bare states or on amplitude pole position. Usually, the
masses of the bare states found in the t-dependant fits are at 20-30 MeV
lower then those obtained in fits without t-dependence.

\subsection{K-matrix two-pole solutions}

The descriptions of the data for the
data-sets A and B are shown in fig.2 (solutions A-1 and B-1).
The K-matrix pole masses,
coupling constants and pole position in the
$K\pi\to K\pi$  $S$-wave scattering amplitude for both solutions
are presented in Table 2.  Both solutions give values of bare
masses and amplitude pole positions which are rather close to each
other.  So we can treat these solutions as  one:
 \be
K_0^{bare}(1220^{+50}_{-60})\, ,\qquad K_0^{bare}(1885^{+50}_{-80})\, .
\label{f1}
\ee

The $K\pi$ scattering amplitude in this case has the poles at
\be
\mbox{II sheet}&\;M=1415 \pm 25-i\;(165 \pm 25)&\;\mbox{MeV}
\nonumber \\
\mbox{III sheet}&\;M=1525 \pm 125-i\;(420 \pm 80)&\;\mbox{MeV}.
\nonumber \\
\mbox{III sheet}&\;M=1820 \pm 40-i\;(125 \pm 50)&\;\mbox{MeV}.
\label{f2}
\ee
The definition of the sheets on the complex $M_{K\pi}$-plane is the
following: II sheet is located under $K\pi$ and $K\pi \pi \pi$
cuts and the sheet III under  $K\pi$,  $K\pi \pi \pi$ and
$K \eta'$ cuts.
The state $K_0^{bare}(1220^{+50}_{-60})$ reveals itself as two poles
of the amplitude in the region $M=1400 - 1550$ MeV, where the $K\eta'$
threshold is located. This situation is analogous to that of
$f_0(980)$: this is also realized as two poles near the $K\bar K$
threshold \cite{f980}.

Classification of the bare scalar states based on eq. (\ref{f1})
is as follows:
\be
~&1^3P_0: & f_0^{bare}(720\pm 100),~f_0^{bare}(1260^{+100}_{-30}),~
K_0^{bare}(1220^{+50}_{-60}) ,~ a_0^{bare}(960\pm 30).~~~
\label{cl1}
\ee
For the nonet of the first radial excitation, following  \cite{km1900},
we have two cases. They are either Solution I:
\be
~&2^3P_0:& f_0^{bare}(1600\pm 50),~f_0^{bare}(1810^{+30}_{-100}),~
K_0^{bare}(1885^{+50}_{-80}) ,~ a_0^{bare}(1640\pm 40),
\nonumber \\
~&\mbox{Glueball:}& f_0^{bare}(1235^{+150}_{-30}),
\label{cl2}
\ee
or Solution II
\be
~&2^3P_0:& f_0^{bare}(1235^{+150}_{-30}),~f_0^{bare}(1810^{+30}_{-100}),~
K_0^{bare}(1885^{+50}_{-80}) ,~ a_0^{bare}(1640\pm 30)
\nonumber \\
~&\mbox{Glueball:}&\qquad f_0^{bare}(1600\pm 50).
\label{cl3}
\ee

Let us stress that the position of
K-matrix poles in the $K \pi$-amplitude
differ by about 100-200 MeV from the
pole positions in scattering amplitudes; this is rather important
from the point of view of the $q\bar q$
classification.

\subsection{K-matrix three-pole solutions}

We have performed two types of three-pole K-matrix fits. In one of them,
the region $M_{\pi K} < 1600$ MeV is described by one pole, while
for the region $M_{\pi K} > 1600$ MeV  two K-matrix poles are used:
these are Solutions A-2 and B-2.

For the next type of solution, we use two K-matrix poles for
the region $M_{\pi K} < 1600$ MeV, while the region $M_{\pi K} > 1600$
MeV is described by one K-matrix pole, Solution B-3.

{\bf Solutions A-2 and B-2}

 Parameters of these solutions are given in Table 3. As before, the
values of the K-matrix bare masses and positions of poles are close to
each other for Solutions A-2 and B-2; thus
\be
K_0^{bare}(1220\pm 70)\, ,\qquad K_0^{bare}(1860\pm 90)\,
\qquad K_0^{bare}(1975\pm 115)\,.
\label{f11}
\ee

The $K\pi$ scattering amplitude in this case has poles at
\be
\mbox{II sheet}&\;M=1420 \pm 30-i\;(170 \pm 30)&\;\mbox{MeV}
\nonumber \\
\mbox{III sheet}&\;M=1530 \pm 125-i\;(400 \pm 150)&\;\mbox{MeV}
\nonumber \\
\mbox{III sheet}&\;M=1815 \pm 40-i\;(100 \pm 50)&\;\mbox{MeV}.
\nonumber \\
\mbox{III sheet}&\;M=2010 \pm 90-i\;(300 \pm 250)&\;\mbox{MeV}.
\label{f22}
\ee

Similarly to the previous solution, the state $K_0^{bare}(1220\pm 70)$
reveals itself as two poles of the amplitude in the region $M=1400 -
1550$ MeV, where the $K\eta'$ threshold is located.

Classification of the bare scalar states based on (\ref{f11})
is as follows:
\be
~&1^3P_0: & f_0^{bare}(720\pm 100),~f_0^{bare}(1260^{+100}_{-30}),~
K_0^{bare}(1220\pm 70) ,~ a_0^{bare}(960\pm 30).~~~
\label{cl11}
\ee
For  the first radial excitation
we have two cases again.  Solution I:
\be
~&2^3P_0:& f_0^{bare}(1600\pm 50),~f_0^{bare}(1810^{+30}_{-100}),~
K_0^{bare}(1860\pm 90) ,~ a_0^{bare}(1640\pm 40),
\nonumber \\
~&\mbox{Glueball:}& f_0^{bare}(1235^{+150}_{-30}).
\label{cl22}
\ee
 Solution II:
\be
~&2^3P_0:& f_0^{bare}(1235^{+150}_{-30}),~f_0^{bare}(1810^{+30}_{-100}),~
K_0^{bare}(1860\pm 90) ,~ a_0^{bare}(1640\pm 30)
\nonumber \\
~&\mbox{Glueball:}&\qquad f_0^{bare}(1600\pm 50).
\label{cl33}
\ee

{\bf Solution B-3}

This solution differs from previous ones and
gives quite different sets of states both for the
$1^3P_0$ nonet  and  for $2^3P_0$.
 It has the following three bare $K_0$
states:
\be
K_0^{bare}(1090\pm 40)\, ,\qquad K_0^{bare}(1375^{+ 125}_{-40})\,
\qquad K_0^{bare}(1950^{+70}_{-20})\,.
\label{f111}
\ee

The $K\pi$ scattering amplitude in this case has poles at
\be
\mbox{II sheet}&\;M=998 \pm 15-i\;(80 \pm 15)&\;\mbox{MeV}
\nonumber \\
\mbox{II sheet}&\;M=1426 \pm 15-i\;(182 \pm 15)&\;\mbox{MeV}
\nonumber \\
\mbox{III sheet}&\;M=1468 \pm 30-i\;(309 \pm 15)&\;\mbox{MeV}
\nonumber \\
\mbox{III sheet}&\;M=1815 \pm 25-i\;(130 \pm 25)&\;\mbox{MeV}.
\label{f222}
\ee

Classification of the bare scalar states based on eq.(\ref{f111})
is as follows:
\be
~&1^3P_0: & f_0^{bare}(720\pm 100),~f_0^{bare}(1260^{+100}_{-30}),~
K_0^{bare}(1090\pm 40) ,~ a_0^{bare}(960\pm 30).~~~
\label{cl111}
\ee
For  the first radial excitation
we have two cases again.  Solution I:
\be
~&2^3P_0:& f_0^{bare}(1600\pm 50),~f_0^{bare}(1810^{+30}_{-100}),~
K_0^{bare}(1375^{+ 125}_{-40}) ,~ a_0^{bare}(1640\pm 40),
\nonumber \\
~&\mbox{Glueball:}& f_0^{bare}(1235^{+150}_{-30}).
\label{cl222}
\ee
 Solution II:
\be
~&2^3P_0:& f_0^{bare}(1235^{+150}_{-30}),~f_0^{bare}(1810^{+30}_{-100}),~
K_0^{bare}(1375^{+ 125}_{-40}) ,~ a_0^{bare}(1640\pm 40)
\nonumber \\
~&\mbox{Glueball:}&\qquad f_0^{bare}(1600\pm 50).
\label{cl333}
\ee
In Solution B-3 the positions of bare states inside nonets are more
compact.

\section{Conclusion}

We have performed the K-matrix analysis of the $\frac120^+$ wave,
thus clarifying the situation with
exotic mesons in the region of $900-1900$ MeV.
Analysis shows that
 there is no basis to consider the resonances $f_0(980)$ and
$a_0(980)$ as superfluous for $q\bar q$ classification:
they are formed from the bare states which are 
members of the $1^3P_0 \; q\bar q$ nonet.
Two variants for
this nonet  are found which differ only in the mass of the lightest
scalar  kaon:
 \be
1^3P_0:  \qquad f_0^{bare}(720\pm
100),~f_0^{bare}(1260^{+100}_{-30}),~ K_0^{bare}(1090\pm 40),~
a_0^{bare}(960\pm 30).~~~ \label{cl1111}
\ee
$$ 1^3P_0: \qquad
f_0^{bare}(720\pm 100),~f_0^{bare}(1260^{+100}_{-30}),~
K_0^{bare}(1220\pm 70),~ a_0^{bare}(960\pm 30).~~~
$$
\newline
 There are more possibilities in construction of the nonet
$2^3P_0 \; q\bar q$: they are given by eqs. (22), (23) and
(33). But in all  schemes, there
exists one extra state, either $f_0^{bare}(1235)$
or $f_0^{bare}(1600)$, which
should be classified as good candidates for the lightest scalar
glueball.  We remind that the state $f_0^{bare}(1600)$, as
a candidate for the lightest glueball, is in agreement with the lattice
gluodynamic calculations \cite{ukqcd}.

\section{Acknowledgement}

We thanks V.V.Anisovich and D.V.Bugg for usefull discussion and remarks.

This work was supported by RFFI grant 96-02-17934 and INTAS-RFBR
grant 95-0267. A.V.A. is grateful for support from INTAS grant
93-0283-ext.

\newpage
\begin{center}
Table 1\\
Couplings $K_0\to$ {\it two mesons} in the leading
and next-to-leading terms
of the $1/N$ expansion. $\Theta$ is the mixing angle for $\eta -
\eta'$ mesons: $\eta=n \bar n \cos \Theta- s \bar s \sin \Theta$
and $\eta'=n \bar n \sin \Theta+ s \bar s \cos \Theta$ where
$n \bar n = (u \bar u +d \bar d)/ \sqrt{2}$.
\vskip 0.5cm
\begin{tabular}{|l|c|c|}
\hline
~ & ~ & ~\\
Channel & Coupling constants & Next-to-leading\\
        & in the leading $1/N_c$ terms & terms \\
~ & ~ & ~\\
\hline
~ & ~ &~ \\
$K^- \pi^+$ & $g^L/2$ & 0 \\
~ & ~ & ~\\
$K^0 \pi^0$ & $g^L/\sqrt{8}$ & 0\\
~ & ~ & ~\\
$K^0 \eta$  & $(\cos \Theta/\sqrt 2 - \sqrt{\lambda}
 \sin \Theta)g^L/2$ &
$\left (\sqrt 2\cos\Theta -\sqrt\lambda\sin\Theta\right)g^{NL}/2$\\
~ & ~& ~ \\
$K^0 \eta'$  & $(\sin\Theta/\sqrt 2 + \sqrt{\lambda}
 \cos \Theta)\;g^L/2$ &
$\left (\sqrt 2\sin\Theta -\sqrt\lambda\cos\Theta\right)g^{NL}/2$\\
~ & ~ & ~\\
\hline
\end{tabular}
\end{center}

\newpage

\begin{center}
Table 2\\
The K-matrix parameters for solutions A-1 and B-1
and position of the poles in the scattering amplitude.
A star denotes that the parameter is fixed.
All values are given in GeV units.
\vskip 0.5cm
\begin{tabular}{|lll|}
\hline
\hline
\multicolumn{3}{|c|}{Solution A-1} \\
\hline
\hline
~ & ~& \\
$M_1=1.234\pm 0.040$ & $g^1_N=1.734^{+0.100}_{-0.200}$ &
                       $g^1_3= 0.139\pm 0.200$ \\
~ & ~ &\\
$M_2=1.870^{+0.040}_{-0.070}$ & $g^2_N=0.741\pm 0.100$ &
                       $g^2_3= 0.363\pm 0.100$ \\
~ & ~ &\\
$f_{11}=0.832\pm 0.20$ & $f_{12}=0.402\pm 0.100$ &
 $f_{13}=0.250\pm 0.100$ \\
~ & $s_0=-1.0^*$ & ~\\
~ & ~ &\\
$a_{3/2}=0.\pm 0.150$ & $f_{3/2}=-1.200\pm 0.200$ &$s_{3/2}= 0^*$  \\
~ & ~& \\
\hline
\multicolumn{3}{|c|}{II sheet poles} \\
$1.427\pm 0.025$     & ~ & ~\\
$-i\;(0.160\pm 0.020)$ & ~ & ~\\
\hline
\multicolumn{3}{|c|}{III sheet poles} \\
$1.525\pm 0.105$     & $1.835\pm 0.030$    & ~ \\
$-i\;(0.370\pm 0.020)$ & $-i\;(0.106\pm 0.020)$ & ~ \\
~ & ~& \\
\hline
\hline
\multicolumn{3}{|c|}{Solution B-1} \\
\hline
\hline
~ & ~ &\\
$M_1=1.202\pm 0.045$ & $g^1_N=1.812^{+0.100}_{-0.200}$ &
                       $g^1_3=-0.048\pm 0.150$ \\
~ & ~ &\\
$M_2=1.900^{+0.040}_{-0.070}$ & $g^2_N=1.035\pm 0.100$ &
                       $g^2_3=0.414\pm 0.100$ \\
~ & ~ &\\
$f_{11}=0.524\pm 0.100$ & $f_{12}=0.226\pm 0.100$
& $f_{13}=-0.449\pm 0.150$ \\
~ & $s_0=-1.0^*$ & ~ \\
~ & ~ &\\
$a_{3/2}=0.\pm 0.150$ & $f_{3/2}=-1.110\pm 0.200$ & $s_{3/2}= 0^*$  \\
~ & ~& \\
\hline
\multicolumn{3}{|c|}{II sheet poles} \\
$1.409\pm 0.015$     & ~ & ~\\
$-i\;(0.185\pm 0.015)$ & ~ & ~\\
\hline
\multicolumn{3}{|c|}{III sheet poles} \\
$1.529\pm 0.125$     & $1.806\pm 0.020$    & ~ \\
$-i\;(0.545\pm 0.050)$ & $-i\;(0.150\pm 0.015)$ & ~ \\
\hline
\end{tabular}
\end{center}
\newpage
\begin{center}
Table 3\\
The K-matrix parameters for solutions A-2 and B-2
and position of the poles in the scattering amplitude.
A star denotes that the parameter is fixed.
All values are given in GeV units.
\vskip 0.1cm
\begin{tabular}{|lll|}
\hline
\hline
\multicolumn{3}{|c|}{Solution A-2} \\
\hline
\hline
~ & ~& \\
$M_1=1.235\pm 0.040$ & $g^1_N=1.739^{+0.100}_{-0.200}$ &
                       $g^1_3= 0.112\pm 0.100$ \\
~ & ~ &\\
$M_2=1.810\pm 0.040$ & $g^2_N=0.454\pm 0.100$ &
                       $g^2_3= 0.391\pm 0.100$ \\
~ & ~ &\\
$M_3=1.947^{+0.070}_{-0.020}$ & $g^3_N=0.493\pm 0.150$ &
                       $g^3_3=-0.368\pm 0.100$ \\
~ & ~ &\\
$f_{11}=0.880\pm 0.20$ & $f_{12}=0.430\pm 0.100$ &
 $f_{13}=0.170\pm 0.100$ \\
~ & $s_0=-1.0^*$ ~ & \\
~ & ~ &\\
$a_{3/2}=0.\pm 0.150$ & $f_{3/2}=-1.214\pm 0.200$ &$s_{3/2}= 0^*$  \\
~ & ~& \\
\hline
\multicolumn{3}{|c|}{II sheet poles} \\
$1.426\pm 0.030$     & ~ & ~\\
$-i\;(0.157\pm 0.025)$ & ~ & ~\\
\hline
\multicolumn{3}{|c|}{III sheet poles} \\
$1.505\pm 0.035$     & $1.795\pm 0.025$     & $1.945\pm 0.025$   \\
$-i\;(0.360\pm 0.015)$ & $-i\;(0.067\pm 0.015)$ & $-i\;(0.068\pm
0.015)$ \\ ~ & ~& \\
\hline
\hline
\multicolumn{3}{|c|}{Solution B-2} \\
\hline
\hline
~ & ~ &\\
$M_1=1.202\pm 0.050$ & $g^1_N=1.902^{+0.100}_{-0.200}$ &
                       $g^1_3=0.239\pm 0.100$ \\
~ & ~ &\\
$M_2=1.912\pm 0.040$     & $g^2_N=0.741\pm 0.200$ &
                       $g^2_3= 0.071\pm 0.150$ \\
~ & ~ &\\
$M_3=1.993\pm 0.100$     & $g^3_N=0.973\pm 0.100$ &
                       $g^3_3= 0.615\pm 0.100$ \\
~ & ~ &\\
$f_{11}=0.648\pm 0.100$ & $f_{12}=0.318\pm 0.100$ &$s_0=-1.0^*$    \\
~ & ~ &\\
$a_{3/2}=0.\pm 0.150$ & $f_{3/2}=-1.018\pm 0.200$ &$s_{3/2}= 0^*$  \\
~ & ~& \\
\hline
\multicolumn{3}{|c|}{II sheet poles} \\
$1.421\pm 0.015$     & ~ & ~\\
$-i\;(0.186\pm 0.015)$ & ~ & ~\\
\hline
\multicolumn{3}{|c|}{III sheet poles} \\
$1.549\pm 0.105$     & $1.836\pm 0.020$     & $2.079\pm 0.030$   \\
$-i\;(0.430\pm 0.125)$ & $-i\;(0.115\pm 0.035)$ & $-i\;(0.475\pm
0.315)$ \\ \hline \end{tabular} \end{center}
\newpage
\begin{center}
Table 4\\
The K-matrix parameters for solution B-3.
and position of the poles in the scattering amplitude.
A star denotes that the parameter is fixed.
All values are given in GeV units.
\vskip 0.5cm
\begin{tabular}{|lll|}
\hline
\hline
\multicolumn{3}{|c|}{Solution B-3} \\
\hline
\hline
~ & ~& \\
$M_1=1.090\pm 0.040$ & $g^1_N=1.545^{+0.200}_{-0.200}$ &
                       $g^1_3= 1.195\pm 0.100$ \\
~ & ~ &\\
$M_2=1.375^{+0.125}_{-0.040}$ & $g^2_N=0.685\pm 0.600$ &
                       $g^2_3=-1.085\pm 0.600$ \\
~ & ~ &\\
$M_3=1.950^{+0.070}_{-0.020}$ & $g^3_N=1.239\pm 0.100$ &
                       $g^3_3= 0.601\pm 0.600$ \\
~ & ~ &\\
$f_{11}=0.176\pm 0.100$ & $f_{12}=0.093\pm 0.100$ &
           $f_{13}=-0.453\pm 0.150$\\
~ & $s_0=-1.0^*$ & ~ \\
~ & ~ &\\
$a_{3/2}=0.\pm 0.150$ & $f_{3/2}=-1.206\pm 0.200$ &$s_{3/2}= 0^*$ \\
~ & ~& \\
\hline
\multicolumn{3}{|c|}{II sheet poles} \\
$0.998\pm 0.015$     & $1.426\pm 0.015$      & ~\\
$-i\;(0.079\pm 0.015)$ & $-i\;(0.182\pm 0.015)$  & ~\\
\hline
\multicolumn{3}{|c|}{III sheet poles} \\
~                 & $1.468\pm 0.050$     & $1.815\pm 0.025$    \\
~                 & $-i\;(0.309\pm 0.015)$ & $-i\;(0.130\pm
0.020)$ \\ \hline \end{tabular} \end{center} \newpage

\newpage
\begin{center}
\epsfig{file=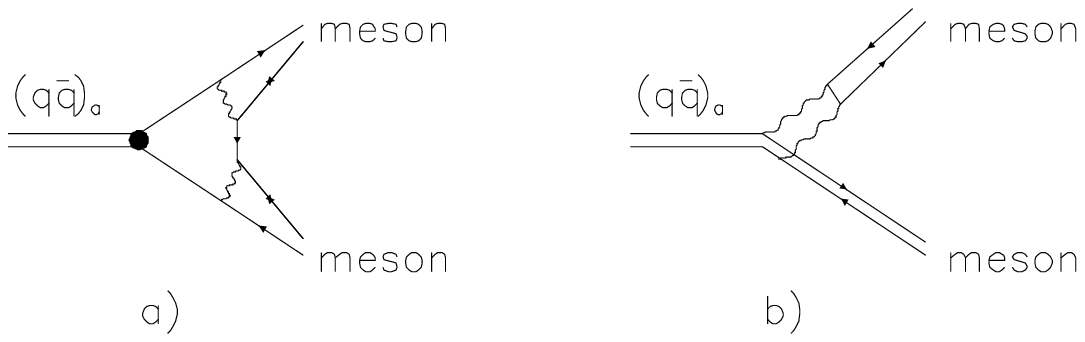,width=16cm}\
Fig. 1. Diagrams of $(q\bar q)_a$-meson decay.
\end{center}
\newpage
\begin{center}
\epsfig{file=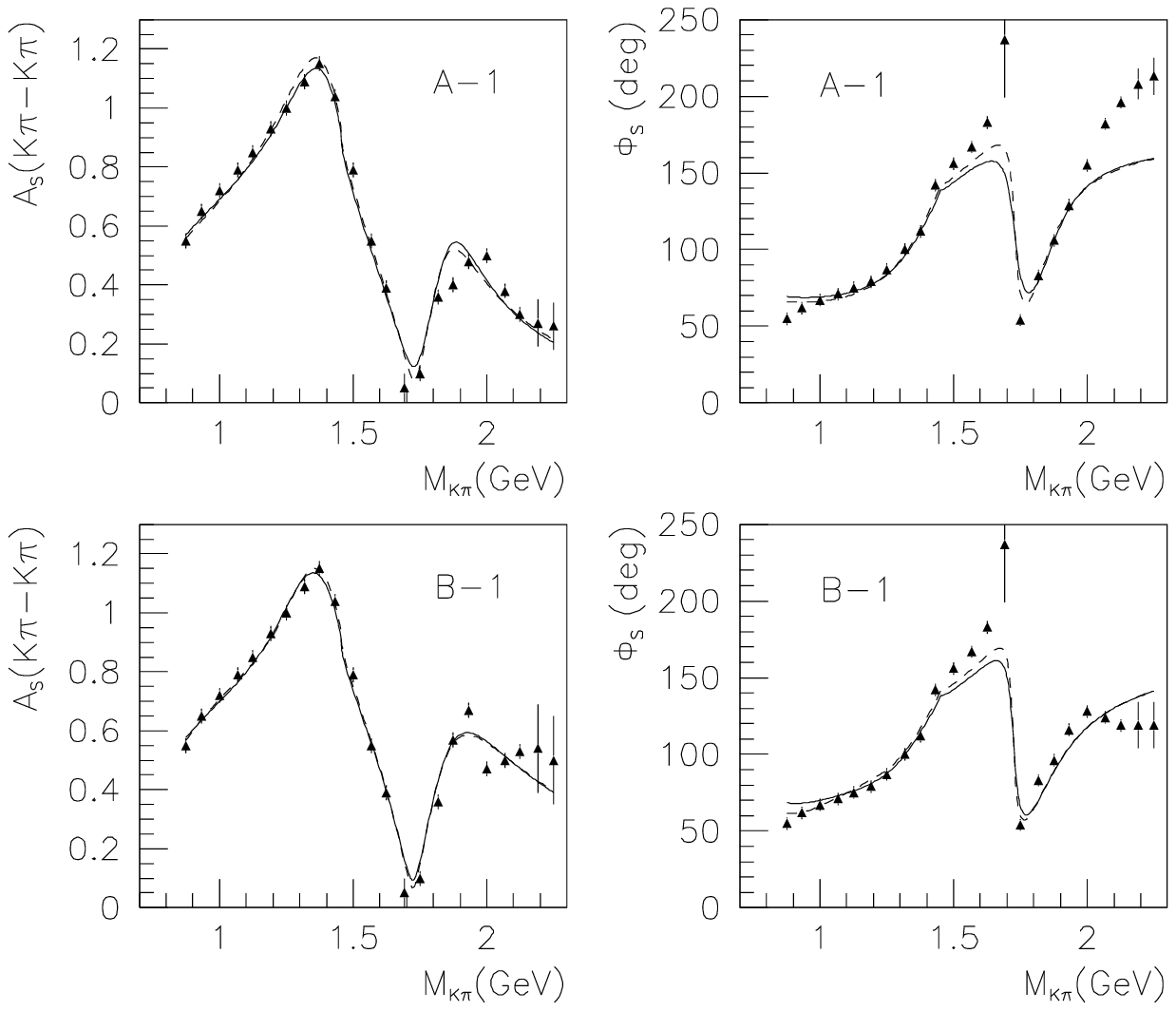,width=16cm}\
Fig. 2. The description of the data sets A
and B from ref. \cite{Aston88}
by the two-pole K-matrix solutions A-1 and B-1.
The solid curves correspond to the fit with
unitar expression for isospin 1/2 amplitude and the dashed ones
to the t-dependant fit.
\end{center}
\newpage
\begin{center}
\epsfig{file=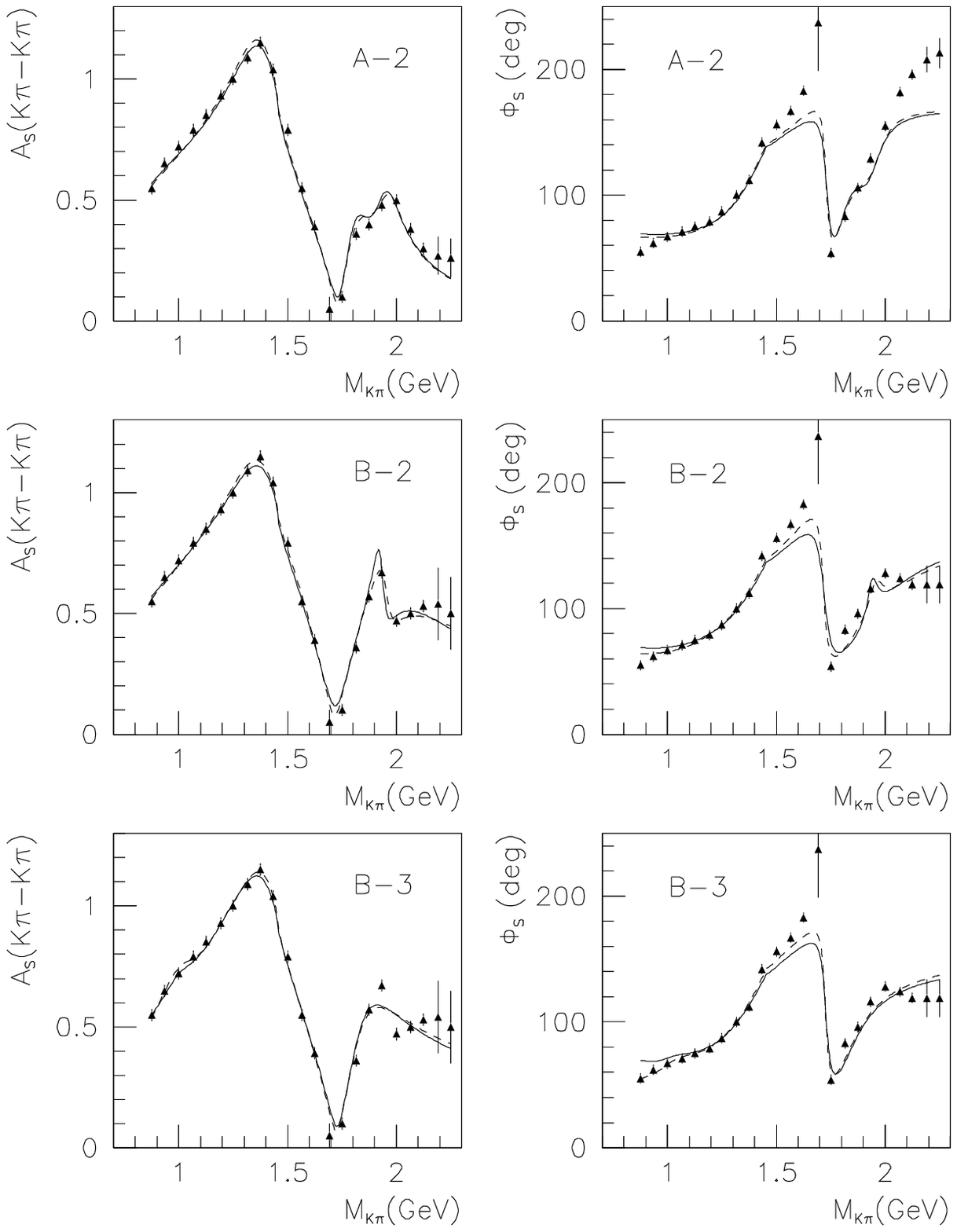,width=16cm}\
Fig. 3. The description of the data of ref. \cite{Aston88}
by three-pole K-matrix solutions A-2, B-2 and B-3. The
definition of curves is the same as in Fig. 2.
\end{center}
\newpage

\end{document}